\documentclass{IEEEtran}

\usepackage{cite}
\usepackage{t1enc}
\usepackage[cmex10]{amsmath}
\usepackage{amssymb}
\usepackage{array}
\usepackage{mdwmath}
\usepackage{mdwtab}
\usepackage[dvipsnames]{xcolor}
\usepackage{pgfplots}

\usepackage{tikz}
\usetikzlibrary{shapes.geometric, arrows}
\usepackage{todonotes} 

\newcommand{\tprev}[1]{#1}

\usetikzlibrary{plotmarks}
\pgfplotsset{compat=newest}
\pgfplotsset{plot coordinates/math parser=false}
\newlength\figureheight
\newlength\figurewidth 
\usepackage{graphicx}
\usepackage{amsmath,amssymb,array,multirow,bigdelim,dsfont}
\usepackage[amssymb]{SIunits} 
\usepackage{url}
\usepackage{t1enc}
\usepackage{url}   
\usepackage{booktabs}
\usepackage{array}
\usepackage[draft]{fixme}
\usepackage{xcolor}
\usepackage[normalem]{ulem}

\def\BibTeX{{\rm B\kern-.05em{\sc i\kern-.025em b}\kern-.08em
    T\kern-.1667em\lower.7ex\hbox{E}\kern-.125emX}}

\begin{document}

\author{Troels~Pedersen and Ramoni Adeogun\thanks{Troels Pedersen and Ramoni Adeogun are with Dept. Electronic Systems, Aalborg University, Denmark. \{troels,ra\}@es.aau.dk. \today. }}

\title{{\fontsize{24}{26}\selectfont{Communication\rule{29.9pc}{0.5pt}}}\break\fontsize{16}{18}\selectfont
Polarimetric Room Electromagnetics}
\author{Troels~Pedersen and Ramoni Adeogun\thanks{Troels Pedersen and Ramoni Adeogun are with Dept. Electronic Systems, Aalborg University, Denmark. \{troels,ra\}@es.aau.dk. \today. }}

\maketitle

 \begin{abstract}
A polarimetric model for the power delay spectrum for inroom communication is proposed.  The proposed model describes the gradual depolarization of the signal with delay. The model is based on the theory of room electromagnetics, specifically the mirror source approach, which is straightforwardly generalized to the polarimetric case. Compared to the previously known unipolarized room electromagnetic models, which are contained as a special case, the new model holds one additional parameter describing the polarization leakage per wall bounce.   \tprev{The proposed model is found to fit well to two sets of polarimetric data one mm-wave and one cm-wave measurements}. 
 \end{abstract}
 \section{Introduction}
 Room electromagnetics \cite{Andersen2007} has attracted interest from many authors  for its ability to characterize the power delay spectrum\tprev{, i.e., mean power delay profile,} of the inroom  radio channel. \tprev{The power delay spectrum is important for the design of communication systems as it determines the mean values of path loss, mean delay, and rms delay spread \cite{Steinboeck2013,Steinbock2015,Pedersen2019, Pedersen2020,Bharti2021}.} Inspired by methods from room acoustics, models have been derived based on a power balance (or Sabine) approach as in \cite{Andersen2007}, or a mirror source (or Eyring) approach as in \cite{Holloway1999,Steinbock2015,Pedersen2018}. Of the two approaches, it was shown in  \cite{Steinbock2015,She2018} that the latter approach  yields more accurate results for the typical values of wall reflection coefficients encountered in room electromagnetics.  The works on room electromagnetics have  almost exclusively focused on uni-polarized propagation for simplicity reason.  However, polarization is commonly considered important to include in channel models \cite{Oestges2008} and indeed, numerous recent geometry- and propagation graph-based stochastic channel models such as \cite{Quitin2010,Gustafson2014,Yin2015,Cheng2017,Karttunen2019,Golmohamadi2020,Adeogun2019} account for polarization effects. 

An exception to the uni-polarized works in room electromagnetics is the  \cite{Cheng2016} which proposed an polarimetric extension for the distance dependent power delay spectrum  model from \cite{Steinboeck2013,Steinbock2015}. This extension was, unfortunately, obtained without actually re-deriving the power delay spectrum from the reverberation theory. The resulting model does not, while being able to fit measurements, link theoretically to the polarimetric propagation mechanisms. Furthermore, the extended model is rather complex in terms of the number of free parameters  to be estimated from measurements. This complicates its use as an analytical tool to study how  propagation scenario affects the polarized channel. 
 
Inroom measurements, such as \cite{Nielsen2011,Gustafson2014,Adeogun2019} suggest that the cross-polar channel, not only show higher power loss, but also exhibit a more gradual onset of the power delay spectrum. \tprev{Although this behaviour affects important parameters, such as the rms delay spread, a gradual onset is  not included in the model \cite{Cheng2016}.} The gradual onset is  predicted well by the polarimetric propagation graph model  \cite{Adeogun2019} for which the polarimetric power delay spectrum was derived by approximating the graph structure, but not derived from  reverberation theory.


    
The present contribution extends the scalar mirror source theory from \cite{Steinbock2015,Pedersen2018,Pedersen2019} to the polarimetric case and derive an analytical closed form expression for the power delay spectrum.  Compared to scalar room electromagnetics, the derived expression contains, apart from antenna characteristics, only one additional propagation parameter $\gamma$  specifying the polarization leakage due to wall interactions.  The power delay spectrum resembles the one from scalar room electromagnetics, with power decay governed by a reverberation time $T$. The polarimetric mixing occurs with a different rate described by the polarimetric mixing time $T_p$ given by the room volume, surface area, and cross-polar leakage $\gamma$. Furthermore,  the cross-polarization ratio (CPR) is also derived for the model. \tprev{To demonstrate the model we fit it to two sets of measurement data from \cite{Gustafson2014,Adeogun2019}. }

 \section{Considered Scenario}
\begin{figure}
    \centering
        \includegraphics[width=0.7\linewidth]{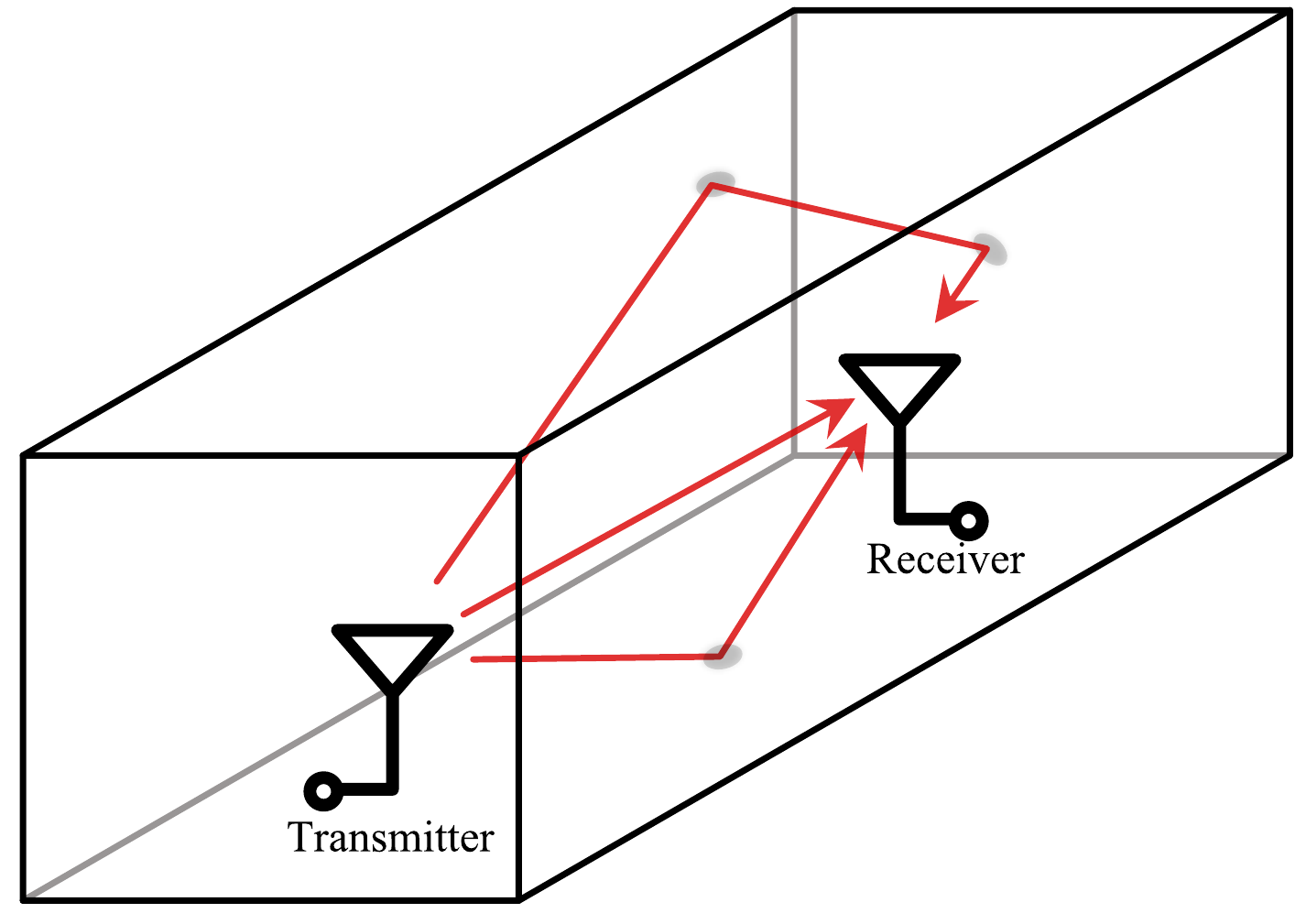}
    \caption{\tprev{Rectangular room with transmitter and  receiver antennas. Here only three of infinitely many propagation paths are shown.}}
    \label{fig:room}
\end{figure}
Consider the propagation of radio waves in a  box-shaped room with a transmitter and a receiver inside \tprev{as the one shown in Fig.~\ref{fig:room}}. Waves emitted by the transmitter bounce back and forth between the walls, ceiling, and floor (hereafter referred to as ``walls'').  The volume of the room is denoted by $V$ and the total
wall area is $S$.  The walls are non-ideal conductors and thus in wall interaction (or bounce), a wave is partially reflected back into the room, attenuated, phase-shifted and possibly with altered polarization state. In each bounce a part of the energy is lost due to absorption in the wall or transmission through it. We assume that the room is isolated in the sense that signals transmitted through walls does not re-enter. Thus we do not distinguish whether signal energy lost due to absorption or transmission.

Directions are specified by the real vector $\Omega$ defined as
$
  \Omega =
[    \cos(\phi)\sin(\theta), 
    \sin(\phi)\sin(\theta),
    \cos(\theta)
 ]^T$
where $\theta$ and $\phi$ are the coelevation and azimuth (in
a right-handed coordinate system with origin at the antenna).  The polarimetric power response of
an antenna in a specific direction $\Omega$ is a two dimensional
vector defined as
$G(\Omega) =
[
  G_\theta(\Omega),
  G_\phi(\Omega)]^T
$ where $G_{\theta/\phi}(\Omega)$ is the power  gain in the $\theta/\phi$ polarization states.
The mean polarimetric gain $\mu$ is defined as an average over the unit sphere $\mathbb S_2$ as 
\begin{equation}
  \label{eq:24}
  \mu =
  \begin{bmatrix}
    \mu_{\theta}\\
    \mu_{\phi}
  \end{bmatrix} = 
  \frac{1}{4\pi}\int_{\mathbb S_2} G(\Omega) d\Omega.
\end{equation}
For a vertically polarized antenna, $\mu_\phi=0$.  Similarly with $\mu_\theta = 0$
we achieve horizontal polarization. 
\tprev{For a lossless antenna, $\mu_\theta + \mu_\phi = 1$  and we may write $\mu = [1-\xi, \xi]^T$ for some constant $0\leq \xi \leq 1$.} In the following, entities related
to the transmitter and receiver are subscripted by  $t$ and $r$, respectively.

Average attenuation and power leakage between polarization states due to a wall bounce is described by a matrix 
\begin{equation}
  \label{eq:17}
  A = g\cdot  M,
\end{equation}
where $0\leq g< 1$ is the average power gain per bounce and the $2\times 2$ matrix $M$ describes the leakage between polarizations.\footnote{\tprev{The two-polarized case is considered here for simplicity and consistency with previous works on indoor polarimetric models. The model can be extended to the tripolarized case, by replacing $M$  by a $3\times 3$ matrix and using the tri-polarized antenna gain pattern in the following derivations. This leads to only minor change in the resulting model, and is omitted here.}} For simplicity, we model $M$ as in \cite{Adeogun2019}, i.e.
\begin{equation}
M  =\frac{1}{1+\gamma}
  \begin{bmatrix}
    1 & \gamma\\
    \gamma & 1
  \end{bmatrix}.
\end{equation}
The cross-polar leakage per bounce is controlled by the parameter  $0\leq\gamma<1$  which varies with the wall materials. With $\gamma=0$, no polarization leakage occurs  whereas for $\gamma\rightarrow 1$, complete depolarization happens after one bounce. 

\section{Power delay Spectrum Model from Mirror Source Theory}\label{sec:PDSModel}
As in \cite{Pedersen2018,Pedersen2019} we consider mirror source analysis which is easily applied to a box-shaped room. Iteratively
mirroring the transmitter in the walls, generates an infinite
collection of mirror sources.   
The signal from each mirror source undergoes successive interactions
(bounces) with the wall. 
 In the following, we first derive the power delay spectrum for case where the transmitter is placed uniformly at random within the room. Next, we introduce the necessary modifications for the case where the transmitter-receiver distance is fixed. 

\subsection{Uniformly Distributed Transmitter Position}
By assuming the transmitter position to  be 
uniformly distributed within the room, the mirror source positions form a homogeneous point process with intensity equal to $1/V$ \cite{Pedersen2018,Pedersen2019}.
Assuming uncorrelated scattering, the power delay spectrum is of the form
\begin{equation}
  \label{eq:19}
  P(\tau) = \mathbb E \left[
    \sum_k  G_r(\Omega_{r,k})^T \frac{A^{B_k} \lambda^2}{4\pi (c\tau_k)^2}
    G_t(\Omega_{t,k}) \delta(\tau-\tau_k)\right ],
\end{equation}
where the mirror sources are indexed by $k$ and the expectation is
taken over the mirror source point process. The speed of light is
denoted by $c$ and the wavelength by $\lambda$. The signal from mirror source $k$ has direction of departure
$\Omega_{t,k}$, direction of arrival $\Omega_{r,k}$ and propagation
delay $\tau_k$. The signal is attenuated due to the inverse squared distance power law and  a number $B_k$ of bounces.  Each bounce attenuates and depolarizes the wave according to the matrix $A$.

The number of bounces can be approximated as a function of
the propagation delay  \cite{Steinbock2015}
\begin{equation}
  \label{eq:20}
  B_k \approx \frac{ cS\tau_k}{4V}. 
\end{equation}
As in \cite{Pedersen2019}, we assume  the mirror source process to be homogeneous and thus the multipath components have uniformly
distributed directions of arrival and departure and arrival rate of  $4\pi u(\tau)c^3\tau^2/V$, where $u(\tau)$ denotes Heaviside's unit step function. Furthermore with good
approximation, the joint intensity function of delays, directions of
departure and arrival factorize into a product of intensity
function, with one factor for delay, and one for each of the directions. 

\begin{figure*}

     \includegraphics[width=\textwidth]{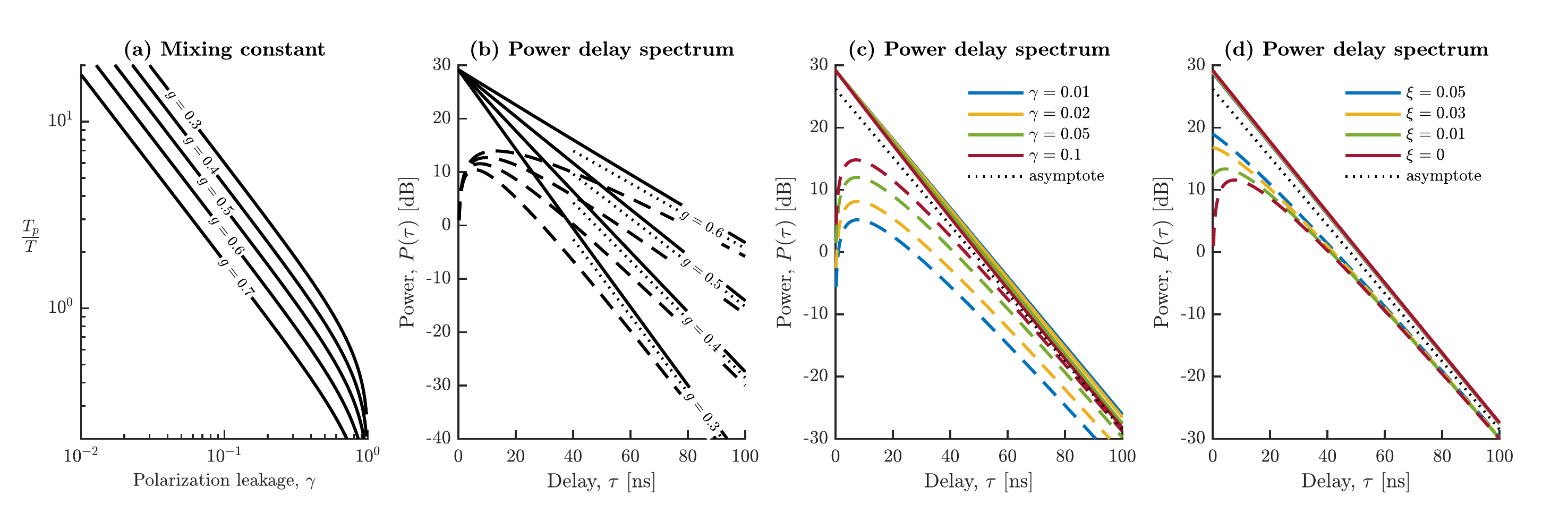}

    \caption{\tprev{Theoretical curves for the proposed model; (a) mixing constant with $g$ as parameter;  (b)--(d) Power delay spectra room with dimensions $3\times4\times 3\,\text{m}^3$. Solid lines: co-polar case with $\mu_t = \mu_r = [1-\xi, \xi]^T$; dashed lines: cross-polar case with  $\mu_t = [1-\xi, \xi]^T$ and  $ \mu_r = [\xi, 1-\xi]^T$. Asymptotes are given in dotted lines. 
One parameter is varied in each plot as stated while the remaining parameters are held fixed at values  $g=0.4$, $\gamma = 0.04$, $\xi =0$.  
Co-polar curves in  (c) and (d)  partly overlap as they vary only a little with the parameter.
}}
    \label{fig:mixingTime}
\end{figure*}

Invoking Campbell's theorem (see e.g. \cite{Jakobsen2014}) and \eqref{eq:24} yields  
\begin{equation}
    P(\tau) =\frac{c \lambda^2 }{V} \mu_r^T A^{cS \tau/4V}\mu_t u(\tau) ,
\end{equation}
Inserting the eigenvalue decomposition $M = Q\Lambda Q^{-1}$  with
\begin{equation}
  \label{eq:18}
\Lambda =
\begin{bmatrix}
  1&0\\
  0& \frac{1-\gamma}{1+\gamma}
\end{bmatrix}
\qquad\text{and}\qquad Q= \frac{1}{\sqrt{2}}
\begin{bmatrix}
  1& 1\\
  1& -1
\end{bmatrix},
\end{equation}
gives after some manipulations 
\begin{multline}
  \label{eq:23}
  P(\tau) = \frac{c\lambda^2 e^{-\tau/T}}{2V} \bigg[\left(\mu_{r,1}\mu_{t,1} +
    \mu_{r,2}\mu_{t,2}\right)
\times \left (1+e^{-\tau/T_p}\right) 
\\
+\left(\mu_{r,1}\mu_{t,2} +
  \mu_{r,2}\mu_{t,1} \right)
\times  \left(1-e^{-\tau/T_p}\right)\bigg]u(\tau).
\end{multline}
Here, $T$ is the usual Eyring reverberation time\footnote{The approximation in \eqref{eq:20} can be improved somewhat by using the Kuttruff approach as described in \cite{Pedersen2018}. This will result in a scaling of both $T$ and $T_p$ by a factor depending on the geometry of the room. This modification  improves the model's prediction accuracy but is omitted here for simplicity. The modification is straightforward to introduce, if greater accuracy is needed.}  \cite{Steinbock2015}  
\begin{equation}
T = -4V\bigg/cS\ln(g).    
\label{eq:29}
\end{equation}
and $T_p$ is a different time constant which we coin the \emph{polarimetric mixing time}
\begin{equation}
  \label{eq:22}
  T_p = -4V \bigg/  cS\ln \left(\frac{1-\gamma}{1+\gamma}\right).  
\end{equation}
\tprev{The value of $\gamma$ affects the  mixing time dramatically:  $\gamma = 0$ gives no polarimetric mixing and $T_p\rightarrow \infty$; $\gamma = 1$ gives $T_p=0$ and the mixing is
instantaneous. Naturally, in realistic scenarios, $\gamma$ lies somewhere in
between these two extremes. \tprev{Fig.~\ref{fig:mixingTime} plots the model \eqref{eq:23} for various parameter settings.} 
}

\tprev{
It appears that both reverberation and mixing time are proportional to  $V/S$.  Thus, both time constants increase with room size. Comparing rooms with fixed volume, flat and elongated rooms have smaller  time constants. We can then define a \emph{mixing constant}
\begin{equation}
    \frac{T_p}{T} = \ln(g)\bigg/\ln \left(\frac{1-\gamma}{1+\gamma}\right),
\end{equation}
which is only dependent on  polarimetric leakage and not on room geometry. Fig.~\ref{fig:mixingTime}(a) reports plots of the mixing constant.}

\tprev{The  model in \eqref{eq:23}  contains  the  classical  unipolarized room electromagnetics model  \cite{Steinbock2015} as special cases. 
With $\gamma =0$, the classical model is recovered;  $\gamma \rightarrow 1$ gives  instantaneous mixing and  results in the classical mode scaled by a constant. In between these two extremes, $P(\tau)$ deviates from the usual exponential decay. However, at large delay it approaches the asymptote (included in Fig.~\ref{fig:mixingTime}) 
 \begin{multline*}
 \frac{c\lambda^2 e^{-\tau/T}}{2V}
 \times
 \left[\left(\mu_{r,1}\mu_{t,1} +
    \mu_{r,2}\mu_{t,2}\right)
+\left(\mu_{r,1}\mu_{t,2} +
  \mu_{r,2}\mu_{t,1} \right)\right],
 \end{multline*}
which equals the classical model up to the bracket term which is unity for lossless antennas.}

\tprev{The power delay spectrum in \eqref{eq:23} can be decomposed into a co-polar term
\begin{equation}
\label{eq:Pco}
P_{\text{co}}(\tau) = 
 \frac{c\lambda^2 e^{-\tau/T}}{2V} \left(\mu_{r,1}\mu_{t,1} +
    \mu_{r,2}\mu_{t,2}\right)
\times \left (1+e^{-\tau/T_p}\right) u(\tau),
\end{equation}
which  has an abrupt onset at $\tau = 0$, and a cross-polar term
\begin{equation}
\label{eq:Pcros}
    P_{\text{cross}}(\tau) = 
    \frac{c\lambda^2 e^{-\tau/T}}{2V}
    \left(\mu_{r,1}\mu_{t,2} +
  \mu_{r,2}\mu_{t,1} \right)
\times  \left(1-e^{-\tau/T_p}\right)u(\tau),
\end{equation}
with a more gradual onset. Their delay-dependent ratio, 
\begin{align}
   \frac{P_{\text{co}}(\tau)}{P_{\text{cross}}(\tau)}
 &= 
  \frac{\left(\mu_{r,1}\mu_{t,1} +
    \mu_{r,2}\mu_{t,2}\right)
}{\left(\mu_{r,1}\mu_{t,2} +
  \mu_{r,2}\mu_{t,1} \right)}
\times
\coth(\tau/2T_p),
  \label{eq:28}
    \end{align}
approaches infinity for small delays and a constant given by the antennas at large delays. The convergence rate of these two terms and thus of the $P(\tau)$ to its asymptote depends on the mixing time.}

The CPR can be computed for the model as
\begin{equation}
    \mathrm{CPR} = \frac{\int_0^\infty  P_{\text{co}}(\tau)d \tau}{\int_0^\infty  P_{\text{cross}}(\tau)d \tau}, \\
 \label{eq:31}
    \end{equation}
which after some straightforward manipulations gives
\begin{equation}
    \mathrm{CPR}=  \frac{\left(\mu_{r,1}\mu_{t,1} +
    \mu_{r,2}\mu_{t,2}\right)
}{\left(\mu_{r,1}\mu_{t,2} +
  \mu_{r,2}\mu_{t,1} \right)}
\times\left(1+2
\frac{T_p}{T}\right).
\end{equation}
\tprev{Thus, the CPR is a function of antenna parameters and the mixing constant.} Considering the extreme cases:
\begin{equation}
    \mathrm{CPR} = 
    \begin{cases}
    \infty & \gamma = 0\\[1ex]
    \displaystyle 
\frac{\left(\mu_{r,1}\mu_{t,1} +
    \mu_{r,2}\mu_{t,2}\right)
}{\left(\mu_{r,1}\mu_{t,2} +
  \mu_{r,2}\mu_{t,1} \right)}    & \gamma \rightarrow 1.
    \end{cases}
\end{equation}


\tprev{For the special case of lossless antennas with perfect cross-polar isolation, the  $\mu$ vectors have zero/one entries only. If both antennas are  vertically polarized,
$\mu_t = \mu_r = [1,0]^T$ we obtain  $P(\tau) = P_{\text{co}}(\tau) $, $P_{\text{cross}}(\tau) = 0$, and  $\mathrm{CPR}$ is infinite; the same results with horizontally polarized antennas, $\mu_t = \mu_r = [1,0]^T$.  
Vertical transmit polarization $\mu_t  = [1,0]^T  $ and horizontal receive polarization $\mu_r  = [0,1]^T$ gives  $P(\tau) = P_{\text{cross}}(\tau) $ and  $P_{\text{co}} =0 $ and  $\mathrm{CPR} = 0$; the same results with $\mu_t  = [0,1]^T  $ and $\mu_r  = [1,0]^T$. We remark that even in the case of unipolarized lossless antennas,  the power delay spectrum in \eqref{eq:23} deviates from the classical exponential decay. }

\subsection{Fixed Transmitter-Receiver Distance}
Up until now, the results were derived assuming the transmitter to be placed uniformly at random within the room, irrespective of the distance to the receiver. Now, we modify the expressions for the case where the transmitter is a fixed distance $d$ away from the receiver.  This corresponds to conditioning on the mirror source point process, so that a sphere of radius $d$ around the receiver is void of any sources. With reasonably good approximation \cite{Pedersen2019}, the conditional intensity of the mirror source process then is $1/V$ everywhere outside this sphere, but zero inside the sphere. The transmitter lie with certainty on the sphere. This gives power delay spectrum conditional on the distance 
\begin{equation}
\label{eq:30}
    P(\tau|d) = P(\tau) u(\tau>d/c) + P_{\mathrm{dir}}(\tau,d),
\end{equation}
where $P_{\mathrm{dir}}(\tau,d)$ describes the direct propagation. If the direct path is completely blocked, $P_{\mathrm{dir}}(\tau,d) = 0$. For the line-of-sight case, with fixed antenna orientation, we have  
\begin{equation}
P_{\mathrm{dir}}(\tau,d) = G_{r}(\Omega_{r,\text{dir}})^T G_{t}(\Omega_{t,\text{dir}})\frac{\lambda^2}{4\pi d^2}\delta(\tau-d/c).
\end{equation}
In the latter expression more parameters are needed as we need to specify the value for the scaling constant $G_{r}(\Omega_{r,\text{dir}})^T G_{t}(\Omega_{t,\text{dir}})$. If this information is available, it can be included into the model. However, we shall for simplicity assume this constant to equal $\mu_r^T\mu_t = \left(\mu_{r,1}\mu_{t,1} +
    \mu_{r,2}\mu_{t,2}\right)$ which corresponds to assuming uniformly distributed orientations. The simplified model reads
\begin{equation}
P_{\mathrm{dir}}(\tau,d) = \left(\mu_{r,1}\mu_{t,1} +
    \mu_{r,2}\mu_{t,2}\right)\frac{\lambda^2}{4\pi d^2}\delta(\tau-d/c).
\end{equation}

We remark  that \eqref{eq:30} is the polarimetric extension of the distance-dependent  spike-plus-exponential model for the power delay spectrum of inroom channels \cite{Steinboeck2013}. Note that
the time constants $T$ and $T_p$ remain unchanged, regardless of the transmitter-receiver distance and antenna characteristics. 

Plugging the distance dependent power delay spectra into \eqref{eq:31} yields
\begin{align}\label{eqCPR}
\mathrm{CPR}(d) =&     
\frac{\left(\mu_{r,1}\mu_{t,1} +
    \mu_{r,2}\mu_{t,2}\right)
}{\left(\mu_{r,1}\mu_{t,2} +
  \mu_{r,2}\mu_{t,1} \right)} \times \\[1ex]
&\qquad \left(Q(d) + \frac{ 
1+ \frac{T_p}{T+T_p}e^{-d/cT_p}}{1 - \frac{T_p}{T+T_p} e^{-d/cT_p}}\right)\notag
\intertext{with the direct propagation term}  
    Q(d)  =&
    \begin{cases}
    0, & \text{in non-line-of-sight}\\
    \frac{\frac{\lambda V}{2\pi c T d^2 }e^{d/cT}}{1-\left(1+\frac{T}{T_p}\right)e^{-d/cT_p}}, &
\text{in line-of-sight.}
    \end{cases}
    \notag
\end{align}

\begin{figure*}
    \centering
    \includegraphics[width=1\textwidth]{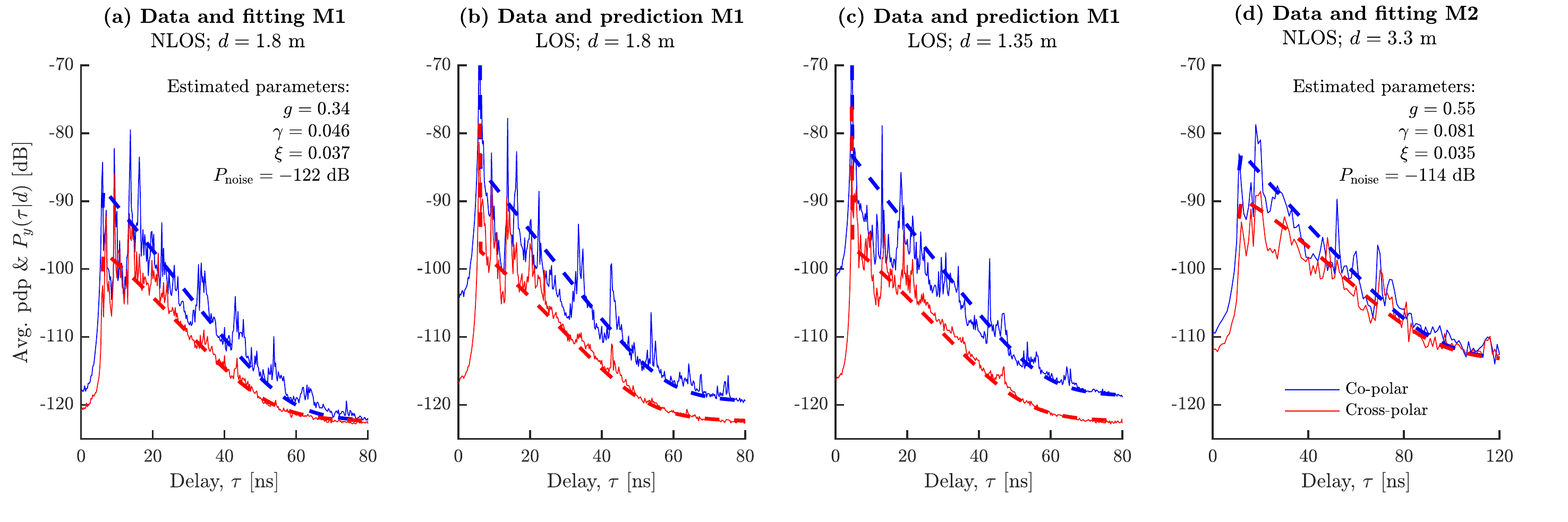}
      \caption{\tprev{Comparison of measured average pdp (solid line) to the power delay spectra model \eqref{eq:noiseAndBandlimited} (dashed line): (a) Data from office room M1, NLOS, $d = 1.8~\text{m}$ with fitted model; (b) Data from M1, LOS, $d = 1.8~\text{m}$ and prediction of model from   (a); (c) Data from M1, LOS, $d = 1.35~\text{m}$  prediction and model prediction of model from (a); 
      (d) Data and model fitting for conference room M2,  NLOS, $d=3.3~\text{m}$. The parameter fitting in (a) gives  $T =6.8~\mathrm{ns}$, $T_p =78~\mathrm{ns}$ and mixing constant of 11.5. For (d)  $T =12.3~\mathrm{ns}$, $T_p =45~\mathrm{ns}$ and mixing constant of 3.7.
      } 
        }
    \label{fig:PDSM}
\end{figure*}

\section{Application to measurement data}

\tprev{We fit the derived model to two sets (M1 and M2) of channel measurements, both obtained with 
a vector network-analyzer and virtual arrays of vertically and horizontally polarized antennas.  The data set M1  from \cite{Gustafson2014} is collected in a furnished meeting room of  $3 \times 4 \times 3~\text{m}^3$ at 
center frequency of $60~\text{GHz}$ with a $4~\text{GHz}$ bandwidth.   A $5\times 5$ virtual planar array with inter-element spacing of 5~mm was used at both the transmitter and receiver. The resulting  $625$ realizations were used to obtain the average power delay profile (pdp) for the co- and cross-polar channels.  Measurements from both line-of-sight (LOS) and non-line-of-sight (NLOS) measurements are included in the data set. The second data set M2 from {\cite{QLiao2017}} is recorded in a $6 \times 10 \times 3~\text{m}^3$ furnished conference room at center frequency  of $15~\text{GHz}$ with a bandwidth of $1~\text{GHz}$. Using a virtual $10\times 10$  array at the transmitter and a single monopole at the receiver, gave  100 realizations per polarization. Further details of M1 and M2 are given  in \cite{Gustafson2014} and \cite{QLiao2017}, respectively.}


\tprev{The model in Section~\ref{sec:PDSModel} characterizes only the propagation effect. The effects of the transmitted signal and measurement noise are included according to 
\begin{equation}
    P_y(\tau|d) = \int P(\tau-t|d)\cdot|s(t)|^2 dt + P_{\mathrm{noise}},
    \label{eq:noiseAndBandlimited}
\end{equation}
where $s(\tau)$ is the transmitted signal  and $P_{\mathrm{noise}}$ is the power of the additive white measurement noise. Information on the antenna responses are unavailable. Thus we set $\mu_t = \mu_r = [1-\xi,\xi]^T$ and $\mu_t  = [1-\xi,\xi]^T; \mu_r = [\xi,1-\xi]^T$ for the co- and cross-polarized channels respectively.}

\tprev{ Parameters $g$ and $\gamma$,  $\xi$ and $P_{\text{noise}}$ are estimated by non-linear least squares fitting  \eqref{eq:noiseAndBandlimited}  (in dB) to the average pdps of the measured co- and cross-polarized channels (also in dB). For M1, the parameters are estimated from one NLOS measurement shown in Fig.~\ref{fig:PDSM}(a) and used to \emph{predict} the LOS measurements in Fig.~\ref{fig:PDSM}(b and c). For M2, the fit is shown in Fig.~\ref{fig:PDSM}(d). The noise power differs between measurements and therefore adapted to each case.}

  \tprev{Fig.~\ref{fig:PDSM}, show clearly that the cross-polar power is present in all four cases. The model fits well both M1 and M2 data as can be seen in Fig.~\ref{fig:PDSM}(a) and (d). Moreover, the predictions shown in Fig.~\ref{fig:PDSM}(b) and (c), are also quite accurate. Thus the model allows for making predictions of the LOS case based on data from an NLOS scenario. 
  We remark that model appears to have slightly different slopes of the co- and cross-polarized power delay spectrum at the plotted delay range. However, it can be verified by setting $ P_{\mathrm{noise}} = 0$ in \eqref{eq:noiseAndBandlimited} that indeed, the two slopes coincide at later delays.  The mixing constant has values 11.5 in M1 and 3.7 in M2. Thus gradual polarimetric mixing occurs in both measurements but occurs faster in M2 than M1. We conjecture that this difference is in part due to the difference in frequency and in part caused by clutter from furniture present in the M2.}  
  

\section{Conclusion}
The polarimetric room electromagnetic model was derived based on reverberation theory to describe the power delay spectrum for an inroom channel. \tprev{We find that depolarization affects the power delay spectrum and thus derived parameters such as rms delay.
 Depolarization occurs gradually over time due to wall interactions, and is specified in the model by a single leakage parameter.} The proposed model is a generalization of the traditional unipolarized room electromagnetic model which is included as a special case.  We introduce the polarimetric mixing time to describe the speed at which depolarization occurs. Remarkably, the ratio of mixing time to reverberation time, termed the mixing constant, depend only on material parameters and not on room geometry.  
\tprev{We fitted the model to data from a 60\ GHz measurement performed in a small meeting room  and a second data set from a 15 GHz measurement obtained  in a larger conference room. 
The model fits well for both data sets,  and was successfully used for prediction of unseen data. The mixing constant was found to be eleven and four for 60~GHz and 15~GHz measurements, respectively. These high values indicate that the gradual depolarization is significant for the considered scenarios. }

\tprev{
With the proposed model we attempt to capture the average depolarization behaviour using a single free parameter. Thus our derivation relies on a simplistic modelling of depolarization due to wall interactions in a clutter-free environment. Despite the simplistic assumption, we found that the model fits well data from real scenarios where clutter is present. The  accuracy can potentially  be improved by using a more detailed model at the price of increased complexity, making the model less suited for analytical work.}

\bibliographystyle{IEEEtran}
\bibliography{references}

\begin{thebibliography}{10}
\providecommand{\url}[1]{#1}
\csname url@samestyle\endcsname
\providecommand{\newblock}{\relax}
\providecommand{\bibinfo}[2]{#2}
\providecommand{\BIBentrySTDinterwordspacing}{\spaceskip=0pt\relax}
\providecommand{\BIBentryALTinterwordstretchfactor}{4}
\providecommand{\BIBentryALTinterwordspacing}{\spaceskip=\fontdimen2\font plus
\BIBentryALTinterwordstretchfactor\fontdimen3\font minus
  \fontdimen4\font\relax}
\providecommand{\BIBforeignlanguage}[2]{{%
\expandafter\ifx\csname l@#1\endcsname\relax
\typeout{** WARNING: IEEEtran.bst: No hyphenation pattern has been}%
\typeout{** loaded for the language `#1'. Using the pattern for}%
\typeout{** the default language instead.}%
\else
\language=\csname l@#1\endcsname
\fi
#2}}
\providecommand{\BIBdecl}{\relax}
\BIBdecl

\bibitem{Andersen2007}
J.~B. Andersen, J.~{\O}. Nielsen, G.~F. Pedersen, G.~Bauch, and J.~M. Herdin,
  ``Room electromagnetics,'' \emph{{IEEE} Antennas Propag. Mag.}, vol.~49,
  no.~2, pp. 27--33, Apr. 2007.

\bibitem{Steinboeck2013}
G.~Steinb{\"o}ck, T.~Pedersen, B.~H. Fleury, W.~Wang, and R.~Raulefs,
  ``Distance dependent model for the delay power spectrum of in-room radio
  channels,'' \emph{{IEEE} Trans. Antennas Propag.}, vol.~61, no.~8, pp.
  4327--4340, Aug. 2013.

\bibitem{Steinbock2015}
------, ``Experimental validation of the reverberation effect in room
  electromagnetics,'' \emph{{IEEE} Trans. Antennas Propag.}, vol.~63, no.~5,
  pp. 2041--2053, May 2015.

\bibitem{Pedersen2019}
T.~Pedersen, ``Stochastic multipath model for the in-room radio channel based
  on room electromagnetics,'' \emph{{IEEE} Trans. Antennas Propag.}, vol.~67,
  no.~4, pp. 2591--2603, Apr. 2019.

\bibitem{Pedersen2020}
------, ``First- and second order characterization of temporal moments of
  stochastic multipath channels,'' in \emph{2020 XXXIIIrd General Assembly and
  Scientific Symposium of the International Union of Radio Science}, 2020, pp.
  1--4.

\bibitem{Bharti2021}
A.~Bharti, R.~Adeogun, X.~Cai, W.~Fan, F.-X. Briol, L.~Clavier, and
  T.~Pedersen, ``Joint modeling of received power, mean delay, and delay spread
  for wideband radio channels,'' \emph{{IEEE} Trans. on Antennas and Propag.},
  pp. 1--1, 2021.

\bibitem{Holloway1999}
C.~Holloway, M.~Cotton, and P.~McKenna, ``A model for predicting the power
  delay profile characteristics inside a room,'' \emph{{IEEE} Trans. Veh.
  Technol.}, vol.~48, no.~4, pp. 1110--1120, Jul. 1999.

\bibitem{Pedersen2018}
T.~Pedersen, ``Modelling of path arrival rate for in-room radio channels with
  directive antennas,'' \emph{{IEEE} Trans. Antennas Propag.}, vol.~66, no.~9,
  pp. 4791--4805, Sep. 2018.

\bibitem{She2018}
J.~She, Y.~Yu, P.-F. Cui, W.-J. Lu, and H.-B. Zhu, ``Reverberation time and
  power model in indoor wireless scenarios,'' \emph{Radioengineering}, vol.~27,
  no.~2, pp. 485--493, Jun. 2018.

\bibitem{Oestges2008}
C.~Oestges, B.~Clerckx, M.~Guillaud, and M.~Debbah, ``Dual-polarized wireless
  communications: from propagation models to system performance evaluation,''
  \emph{{IEEE} Trans. Wireless Commun.}, vol.~7, no.~10, pp. 4019--4031, 2008.

\bibitem{Quitin2010}
F.~Quitin, C.~Oestges, F.~Horlin, and P.~D. Doncker, ``Polarization
  measurements and modeling in indoor {NLOS} environments,'' \emph{{IEEE}
  Trans. Wireless Commun.}, vol.~9, no.~1, pp. 21--25, Jan. 2010.

\bibitem{Gustafson2014}
C.~Gustafson, K.~Haneda, S.~Wyne, and F.~Tufvesson, ``On mm-wave multipath
  clustering and channel modeling,'' \emph{{IEEE} Trans. Antennas Propag.},
  vol.~62, no.~3, pp. 1445--1455, 2014.

\bibitem{Yin2015}
X.~Yin, Y.~He, C.~Ling, L.~Tian, and X.~Cheng, ``Empirical stochastic modeling
  of multipath polarizations in indoor propagation scenarios,'' \emph{{IEEE}
  Trans. Antennas Propag.}, vol.~63, no.~12, pp. 5799--5811, Dec. 2015.

\bibitem{Cheng2017}
X.~Cheng, Y.~He, and M.~Guizani, ``3-d geometrical model for multi-polarized
  {MIMO} systems,'' \emph{{IEEE} Access}, vol.~5, pp. 11\,974--11\,984, 2017.

\bibitem{Karttunen2019}
A.~Karttunen, J.~Jarvelainen, S.~L.~H. Nguyen, and K.~Haneda, ``Modeling the
  multipath cross-polarization ratio for 5{\textendash}80-{GHz} radio links,''
  \emph{{IEEE} Trans. Wireless Commun.}, vol.~18, no.~10, pp. 4768--4778, Oct.
  2019.

\bibitem{Golmohamadi2020}
M.~Golmohamadi and J.~Frolik, ``A geometric scattering model for circularly
  polarized indoor channels,'' \emph{{IEEE} Trans. Antennas Propag.}, vol.~68,
  no.~3, pp. 2290--2296, Mar. 2020.

\bibitem{Adeogun2019}
R.~Adeogun, T.~Pedersen, C.~Gustafson, and F.~Tufvesson, ``Polarimetric
  wireless indoor channel modeling based on propagation graph,'' \emph{{IEEE}
  Trans. Antennas Propag.}, vol.~67, no.~10, pp. 6585--6595, Oct. 2019.

\bibitem{Cheng2016}
S.~Cheng, D.~P. Gaillot, E.~Tanghe, P.~Laly, T.~Demol, W.~Joseph, L.~Martens,
  and M.~Lienard, ``Polarimetric distance-dependent models for large hall
  scenarios,'' \emph{{IEEE} Trans. Antennas Propag.}, vol.~64, no.~5, pp.
  1907--1917, May 2016.

\bibitem{Nielsen2011}
J.~{\O}. Nielsen, J.~B. Andersen, G.~F. Pedersen, and M.~Pelosi, ``On
  polarization and frequency dependence of diffuse indoor propagation,'' in
  \emph{2011 {IEEE} Veh. Techn. Conf. ({VTC} Fall)}.\hskip 1em plus 0.5em minus
  0.4em\relax {IEEE}, Sep. 2011.

\bibitem{Jakobsen2014}
M.~L. Jakobsen, T.~Pedersen, and B.~H. Fleury, ``Analysis of stochastic radio
  channels with temporal birth-death dynamics: A marked spatial point process
  perspective,'' \emph{{IEEE} Trans. Antennas Propag.}, vol.~62, no.~7, pp.
  3761--3775, Jul. 2014.

\bibitem{QLiao2017}
Q.~Liao, Z.~Ying, and C.~Gustafson, ``Simulations and measurements of 15 and 28
  {GHz} indoor channels with different array configurations,'' in \emph{Int.
  Workshop on Antenna Technol.: Small Antennas, Innovative Structures, and
  Applications ({iWAT})}.\hskip 1em plus 0.5em minus 0.4em\relax {IEEE}, 2017,
  pp. 256--259.

\end{thebibliography}
\end{document}